\documentclass[preprint]{aastex}
\usepackage{psfig}
\slugcomment{KSUPT-02/4 \hspace{0.5truecm} August 2002}

\newcommand{\wisk}[1]{\ifmmode{#1}\else{$#1$}\fi}

\begin{document}

\title{OVRO CMB Anisotropy Measurement Constraints on 
       Flat-$\Lambda$ and Open CDM Cosmogonies}

\author{
  Pia~Mukherjee\altaffilmark{1},
  Tarun~Souradeep\altaffilmark{1,2},
  Bharat~Ratra\altaffilmark{1},   
  Naoshi~Sugiyama\altaffilmark{3,4},
  and
  Krzysztof~M.~G\'orski\altaffilmark{5,6}
  }

\altaffiltext{1}{Department of Physics, Kansas State University,
                 116 Cardwell Hall, Manhattan, KS 66506.}
\altaffiltext{2}{Current address: IUCAA, Post Bag 4, Ganeshkhind, Pune 
                 411007, India.}
\altaffiltext{3}{Division of Theoretical Astrophysics, National Astronomical
                 Observatory, 2-21-1 Osawa, Mitaka, Tokyo 181-8588, Japan.}
\altaffiltext{4}{Max Planck Institute for Astrophysics, Karl-Schwarzschild 
                 Str. 1, Postfach 1317, Garching D-85741, Germany.}
\altaffiltext{5}{European Southern Observatory, Karl-Schwarzschild Str. 2, 
                 Garching D-85748, Germany.}
\altaffiltext{6}{Warsaw University Observatory, Aleje Ujazdowskie 4, 
                 00-478 Warszawa, Poland.}

\begin{abstract}
We use Owens Valley Radio Observatory (OVRO) cosmic microwave background 
(CMB) anisotropy data to constrain cosmological parameters. We account for 
the OVRO beamwidth and calibration uncertainties, as well as the 
uncertainty induced by the removal of non-CMB foreground contamination.
We consider open and spatially-flat-$\Lambda$ cold dark matter cosmogonies,
with nonrelativistic-mass density parameter $\Omega_0$ in the range 0.1--1, 
baryonic-mass density parameter $\Omega_B$ in the range (0.005--0.029)$h^{-2}$, 
and age of the universe $t_0$ in the range (10--20) Gyr. Marginalizing over 
all parameters but $\Omega_0$, the OVRO data favors an open
(spatially-flat-$\Lambda$) model with $\Omega_0\simeq$ 0.33 (0.1). At the 
2 $\sigma$ confidence level model normalizations deduced from the OVRO 
data are mostly consistent with those deduced from the DMR, UCSB South 
Pole 1994, Python I-III, ARGO, MAX 4 and 5, White Dish, and SuZIE data sets. 
\end{abstract}


\section{Introduction}

Cosmic microwave background (CMB) anisotropy measurements have begun to
provide interesting constraints on cosmological parameters.\footnote{
See, e.g., Miller et al. (2002a), Coble et al. (2001), Scott et al. (2002),
and Mason et al. (2002) for recent measurements, and, e.g., Podariu et al. 
(2001), Wang, Tegmark, \& Zaldarriaga (2002), Durrer, Novosyadlyj,
\& Apunevych (2001), and Miller et al. (2002b) for recent discussions of
constraints on cosmological parameters.}
Ganga et al. (1997a, hereafter GRGS) developed a technique to account for 
uncertainties, such as those in the beamwidth and the calibration, in 
likelihood analyses of CMB anisotropy data. This technique has been used 
with theoretically-predicted CMB anisotropy spectra in analyses of the 
Gundersen et al. (1995) UCSB South Pole 1994 data, the Church et al. (1997)
SuZIE data, the Lim et al. (1996) MAX 4+5 data, the Tucker et al. (1993) 
White Dish data, the de Bernardis et al. (1994) ARGO data, and the Platt 
et al. (1997) Python I-III data (GRGS; Ganga et al. 1997b, 1998; Ratra 
et al. 1998, 1999a, hereafter R99a; Rocha et al. 1999, hereafter R99). 
A combined analysis of all these data sets, excluding the Python data, is 
presented in Ratra et al. (1999b, hereafter R99b).

In this paper we present a similar analysis of CMB anisotropy data from the 
OVRO observations (Leitch et al. 2000, hereafter L00). The OVRO detectors 
and telescopes are described in Leitch (1998) and L00; here we review 
information about the experiment that is needed for our analysis.

OVRO data were taken in two frequency bands, one centered at 14.5 GHz 
(Ku band), the other at 31.7 GHz (Ka band). Thirty-six fields, along an
approximate circle at declination $\delta \simeq 88^\circ$ centered on 
the North Celestial Pole (NCP) were observed. In our computations we
use the coordinates for the 36 fields given in Table 2 of L00. The OVRO 
measurements were made by switching the beam in a two-point pattern along the 
circle, resulting in a three-beam response to the sky signal. The beamthrow 
is $22'{\!}.16$. The zero-lag window function parameters for the OVRO 
experiment are given in Table 1. This and other window functions are 
shown in Fig. 18 of L00.

L00 use multiepoch VLA observations to detect and remove non-CMB discrete 
source contamination from the OVRO data. We have also analyzed the OVRO data
ignoring 3 of the 36 fields that were affected by the strongest variable 
discrete source; cosmological constraints derived from this restricted 
OVRO CMB anisotropy data set are very consistent with those derived from
the full OVRO CMB anisotropy data set, so we do not discuss this restricted 
OVRO data set analysis further. 

Since OVRO data were taken at two frequencies, it is possible to fit the
data to both a non-CMB foreground component (parametrized by the frequency
dependent temperature anisotropy $\Delta T_{\rm fore} \propto \nu^\beta$)
and a CMB anisotropy component with spectral index $\beta = 0$.\footnote{
See L00 and Mukherjee et al. (2002) for discussions of foreground contaminants
in the OVRO microwave data.}
We use the method in $\S$ 11 of L00 to extract the CMB anisotropy component 
in the OVRO data, marginalizing over a foreground spectral index in the 
range $-3 < \beta < 2$ in our likelihood analysis.\footnote{
Although the data themselves are unable to rule out more negative values
of beta (L00, Fig. 14), Leitch et al. (1997) use low frequency maps of the
NCP region to rule out such values.}${^,}$\footnote{
Following Mukherjee et al. (2002) we have also analyzed the 31.7 GHz
OVRO CMB anisotropy data while marginalizing over possible 100 $\mu$m
and 12 $\mu$m foreground contaminant template (Schlegel, Finkbeiner, 
\& Davis 1998) correlated components. The cosmological constraints from 
these analyzes are quite consistent with results presented here. This is 
because although the foreground signal inferred in our analysis is not 
entirely fit by the dust data, they are significantly correlated, and the 
31.7 GHz data, modelled either way, is almost entirely CMB anisotropy.
The OVRO data at its two frequencies are shown in Fig. 13 of L00, 
the deduced CMB anisotropy and foreground signals are shown in Fig. 16 
of L00, and the dust-correlated emission is shown in Fig. 1 of Mukherjee
et al. (2002).}

CMB anisotropy constraints are derived from the foreground-corrected
31.7 GHz data.\footnote{
At $\beta = -2.2$ for the foreground contaminant, $96\%$ of the 31.7 GHz 
data is CMB anisotropy.}
The 31.7 GHz beam profile is well approximated by a circular Gaussian
of FWHM $7'{\!}.37 \pm 0'{\!}.26$ (one standard deviation uncertainty).
We use the method of GRGS to account for the OVRO beam uncertainty. 

As discussed in L00, the noise in the 31.7 GHz data indicates the
presence of a component that is correlated between neighboring fields
(this component is small compared to the uncorrelated noise in a single
scan of data). As a result the 31.7 GHz OVRO data show only one-half of
the anticorrelation for nearest neighbor fields expected for a triple
beam chopped experiment.\footnote{
Models that neglect these correlations are grossly discrepant with the 
data, while when these correlations are accounted for the model fits are 
reasonable and consistent with the data. This can be seen from Fig. 19 
of L00 and we find the same.} 
This one-offdiagonal correlated noise is included and its amplitude 
marginalized over in our analysis.

A constant offset is removed from the OVRO data; we marginalize over the
amplitude of the offset to account for this in our likelihood analysis. 
The 1 $\sigma$ absolute calibration uncertainty of the OVRO data is $4.3\%$,
and the method developed by GRGS is used to account for it. 

In $\S$ 2 we summarize the computational techniques used in our analysis. See 
GRGS and R99a for detailed discussions. Results are presented and discussed 
in $\S$ 3. We conclude in $\S$4.

\section{Summary of Computation}

In this paper we focus on a spatially-flat CDM model with a cosmological 
constant $\Lambda$.\footnote{
See, e.g., Peebles (1984), Efstathiou, Sutherland, \& Maddox (1990), 
Stompor, G\'orski, \& Banday (1995), Ratra et al. (1997), Sahni \& 
Starobinsky (2000), Carroll (2001), and Peebles \& Ratra (2002). 
While not considered in this paper, a time-variable dark energy dominated
spatially-flat model is also largely consistent with current observations
(see, e.g., Peebles \& Ratra 1988; Ratra \& Quillen 1992; Steinhardt 1999;
Brax, Martin, \& Riazuelo 2000; Huterer \& Turner 2001; Chen \& Ratra 2002;
Deustua et al. 2002).}
As a foil we also consider a spatially open model with no $\Lambda$ (see,
e.g., Gott 1982; Ratra \& Peebles 1995). These models are discussed in
more detail in R99a, R99b, and R99.

The CMB anisotropy spectra in these models are generated from quantum 
fluctuations in weakly coupled fields during an early epoch of inflation
and so are Gausssian (see, e.g., Ratra 1985; Fischler, Ratra, \& Susskind
1985). Consistent with this, the observed smaller-scale CMB anisotropy 
appears to be Gaussian (see, e.g., Park et al. 2001, Wu et al. 2001;
Shandarin et al. 2002; Polenta et al. 2002), and the experimental noise also 
appears to be Gaussian, thus validating our use of the GRGS likelihood 
analysis method. 

As discussed in R99a, the spectra are parameterized by their quadrupole-moment
amplitude $Q_{\rm rms-PS}$, the nonrelativistic-mass density parameter 
$\Omega_0$, the baryonic-mass density parameter $\Omega_B$, and the age of the 
universe $t_0$. The spectra are computed for a range of $\Omega_0$ spanning the 
interval 0.1 to 1 in steps of 0.1, for a range of $\Omega_B h^2$ [the Hubble 
parameter $h = H_0/(100\ {\rm km}\ {\rm s}^{-1}\ {\rm Mpc}^{-1})$] spanning 
the interval 0.005 to 0.029 in steps of 0.004, and for a range of $t_0$ 
spanning the interval 10 to 20 Gyr in steps of 2 Gyr. In total 798 spectra were 
computed to cover the cosmological-parameter spaces of the open and 
flat-$\Lambda$ models. Examples of spectra are shown in Fig. 2 of R99a,
Fig. 1 of R99b, and Fig. 2 of R99.

Following GRGS, for each of the 798 spectra considered the ``bare" likelihood 
function is computed at the nominal beamwidth and calibration, as well as at 
a number of other values of the beamwidth and calibration determined from the 
measurement uncertainties. The likelihood function used in the derivation of 
the central values and limits is determined by integrating (marginalizing) the 
bare likelihood function over the beamwidth and calibration uncertainties with
weights determined by the measured probability distribution functions of the
beamwidth and the calibration. See GRGS for a more detailed discussion. The 
likelihoods are a function of four parameters mentioned above: 
$Q_{\rm rms-PS}$, $\Omega_0$, $\Omega_B h^2$, and $t_0$. We also compute 
marginalized likelihood functions by integrating over one or more of these 
parameters after assuming a uniform prior in the relevant parameters.
The prior is set to zero outside the ranges considered for the parameters.
GRGS and R99a describe the prescription used to determine central values and 
limits from the likelihood functions. In what follows we consider 1, 2, and
3 $\sigma$ highest posterior density limits which include 68.3, 95.4, and
99.7\% of the area.

\section{Results and Discussion}

Table 2 lists the derived values of $Q_{\rm rms-PS}$ and bandtemperature 
$\delta T_l$ for the flat bandpower spectrum, for the OVRO data. These 
numerical values account for the correlated noise and offset removal, the 
beamwidth and 
calibration uncertainties, and the uncertainty due to non-CMB diffuse 
foreground contamination removal. These results are very consistent with 
those of L00. For the flat bandpower spectrum the OVRO data average 1 
$\sigma$ $\delta T_l$ error bar is $\sim 14\%$\footnote{
For comparison, the corresponding 1 $\sigma$ $\delta T_l$ error bar is 
$\sim 10-12\%$ for DMR (depending on model, G\'orski et al. 1998), 
$\sim 15\%$ for ARGO (R99a), and $\sim 14\%$ for MAX 4+5 (Ganga et al. 1998).} 
: OVRO data results in a very significant detection of CMB anisotropy, even 
after accounting for the uncertainties listed above. 

As discussed in R99a, R99b, and R99, the four-dimensional posterior probability 
density distribution function $L(Q_{\rm rms-PS}, \Omega_0, \Omega_B h^2, t_0)$ 
is nicely peaked in the $Q_{\rm rms-PS}$ direction but fairly flat in the other 
three directions. Marginalizing over $Q_{\rm rms-PS}$ results in a 
three-dimensional posterior distribution $L(\Omega_0,  \Omega_B h^2, t_0)$ 
which is steeper, but still relatively flat. As a consequence, limits 
derived from the four- and three-dimensional posterior distributions are
generally not highly statistically significant. We therefore do not show 
contour plots of these functions here. Marginalizing over $Q_{\rm rms-PS}$ and 
one other parameter results in two-dimensional posterior probability 
distributions which are more peaked (see Figs. 1). As in the ARGO (R99a), 
Python (R99), and combination (R99b) data set analyses, in some cases these
peaks are at an edge of the parameter range considered.

Figure 1 shows that the two-dimensional posterior distributions allow one to 
distinguish between different regions of parameter space at a fairly high 
formal level of confidence.\footnote{
See Fig. 4 of R99a, Fig. 2 of R99b, and Fig. 3 of R99 for related 
cosmological constraints from other data.}
For instance, the open model near $\Omega_0 \sim 0.75$, $\Omega_B h^2 
\sim 0.03$, and $t_0 \sim 20$ Gyr, and the flat-$\Lambda$ model near 
$\Omega_0 \sim 0.6$, $\Omega_B h^2 \sim 0.03$, and $t_0 \sim 20$ Gyr,
are both formally ruled out at $\sim 3$ $\sigma$ confidence. However, we 
emphasize, as discussed in R99a, R99b, and R99, care must be exercised when 
interpreting the discriminative power of these formal limits, 
since they depend sensitively on the fact that the uniform prior has been set 
to zero outside the range of the parameter space we have considered.

Figure 2 shows the contours of the two-dimensional posterior distribution 
for $Q_{\rm rms-PS}$ and $\Omega_0$, derived by marginalizing the 
four-dimensional distribution over $\Omega_B h^2$ and $t_0$. These are shown
for the OVRO and DMR data, for both the open and flat-$\Lambda$ models. Constraints on these parameters from the OVRO data are consistent with those 
from the DMR data.

Figure 3 shows the one-dimensional posterior distribution functions for 
$\Omega_0$, $\Omega_B h^2$, $t_0$, and $Q_{\rm rms-PS}$, derived by 
marginalizing the four-dimensional posterior distribution over the other three 
parameters. From these one-dimensional distributions, the OVRO data favors 
an open (flat-$\Lambda$) model with $\Omega_0$ = 0.33 (0.10), or 
$\Omega_B h^2$ = 0.005 (0.005), or $t_0$ = 10 (11) Gyr, amongst the models
considered. At 2 $\sigma$ confidence the OVRO data formally rule out 
only small regions of parameter space. From the one-dimensional distributions
of Fig. 3, the data require $\Omega_0$ $< 0.69$ or $> 0.75$
($\Omega_0$ $< 0.51$ or $> 0.56$), or $\Omega_B h^2$ $< 0.028$ 
($\Omega_B h^2$ $< 0.028$), or $t_0$ $< 19$ Gyr ($t_0$ $< 19$ Gyr) for the 
open (flat-$\Lambda$) model at 2 $\sigma$.

While the statistical significance of the constraints on cosmological 
parameters is not high, it is reassuring that the OVRO data favor 
low-density, young models, consistent with indications from most other
data. The constraints on $\Omega_B h^2$ derived from the OVRO data are
somewhat puzzling. They are more consistent with those derived from the
Python and combination CMB anisotropy data sets analyzed by R99 and R99b,
but less so with those from ARGO (R99a) and more recent data sets 
(Netterfield et al. 2002; Pryke et al. 2002; Stompor et al. 2001) 
which favor higher $\Omega_B h^2$.
The lower $\Omega_B h^2$ found here is more consistent with the low 
Cyburt, Fields, \& Olive (2001) standard nucleosynthesis value determined
from helium and lithium abundance measurements, and less consistent with the
high deuterium-based value of Burles, Nollett, \& Turner (2001). 

The peak values of the one-dimensional posterior distributions shown in 
Fig. 3 are listed in the figure caption for the case when the 
four-dimensional posterior distributions are normalized such that
$L(Q_{\rm rms-PS}\ =\ 0\ \mu{\rm K})\ =\ 1$. With this normalization, 
marginalizing over the remaining parameter the fully marginalized
posterior distributions are $1.4\times 10^{67}(1.3\times 10^{67})$ for the 
open (flat-$\Lambda$) model. This is not
inconsistent with the indication from panels $a)$ and $b)$ of Fig. 3 that the 
most-favored open model is marginally more favored than the most-favored
flat-$\Lambda$ one.

\section{Conclusion}

The OVRO data results derived here are mostly consistent with 
those derived from the DMR, SP94, Python I-III, ARGO, MAX 4+5, White Dish and 
SuZIE data. The OVRO data significantly constrains $Q_{\rm 
rms-PS}$ (for the flat bandpower spectrum $Q_{\rm rms-PS}\ = \ 38{+6 \atop
-5}\ \mu$K at 1 $\sigma$) and weakly favors low-density, low $\Omega_B h^2$, young models.

\bigskip

We acknowledge valuable assistance from R. Stompor and helpful discussions 
with K. Ganga and E. Leitch. PM, BR, and TS acknowledge support 
from NSF CAREER grant AST-9875031. NS acknowledges support from the
Alexander von Humboldt Foundation and Japanese Grant-in-Aid for 
Science Research Fund No. 14540290.

\clearpage

\begin{table}
\begin{center}
\caption{Numerical Values for the Zero-Lag Window Function 
Parameters\tablenotemark{a}}
\vspace{0.3truecm}
\tablenotetext{{\rm a}}{The value of $l$ where $W_l$ is
largest, $l_{\rm m}$, the two values of $l$ where $W_{l_{e^{-0.5}}} =
e^{-0.5} W_{l_{\rm m}}$, $l_{e^{-0.5}}$, the effective multipole,
$l_{\rm e} = I(lW_l)/I(W_l)$, and 
$I(W_l) = \sum^\infty_{l=2}(l+0.5)W_l/\{l(l+1)\}$.}
\begin{tabular}{ccccc}
\tableline\tableline
    $l_{e^{-0.5}}$ & $l_{\rm e}$ & $l_{\rm m}$
  & $l_{e^{-0.5}}$ & $\sqrt{I(W_l)}$  \\
\tableline
 360 &  596 &  537 &  753 & 1.41  \\
\tableline
\end{tabular}
\end{center}
\end{table}


\begin{table}
\begin{center}
\caption{Numerical Values for $Q_{\rm rms-PS}$ and $\delta T_l$ from Likelihood
Analyses Assuming a Flat Bandpower Spectrum}
\vspace{0.3truecm}
\tablenotetext{{\rm a}}{The first of the three entries is 
where the posterior probability density distribution function peaks and the 
vertical pair of numbers are the $\pm 1$ $\sigma$ (68.3\% highest posterior
density) values.} 
\tablenotetext{{\rm b}}{Average absolute error on $Q_{\rm rms-PS}$ in 
$\mu$K.}
\tablenotetext{{\rm c}}{Average fractional error, as a fraction of the 
central value.}
\tablenotetext{{\rm d}}{Likelihood ratio.}
\begin{tabular}{ccccc}
\tableline\tableline
  $Q_{\rm rms-PS}$\tablenotemark{a} & Ave. Abs. Err.\tablenotemark{b} 
& Ave. Frac. Err.\tablenotemark{c} & $\delta T_l$\tablenotemark{a} 
& LR\tablenotemark{d} \\
  ($\mu$K) & ($\mu$K) & {\ } & ($\mu$K) & {\ } \\
\tableline
\medskip
  38 ${44 \atop 33}$ & 5.5 & 14\% & 59 ${69 \atop 51}$ & $9 \times 10^{66}$ \\
\tableline
\end{tabular}
\end{center}
\end{table}

\clearpage


\begin{figure}[p]
\psfig{file=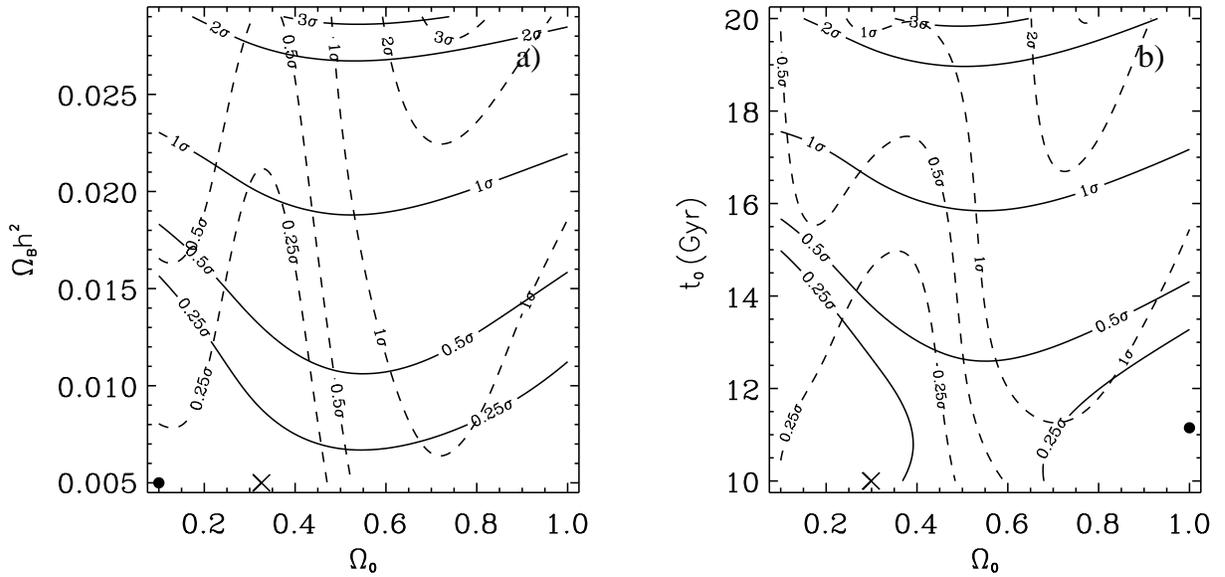,width=6.7in,angle=0}
\caption{Confidence contours and maxima of the OVRO data two-dimensional 
posterior probability density distribution functions, as a function of the 
two parameters on the axes of each panel (derived by marginalizing the 
four-dimensional posterior distributions over the other two parameters). 
Dashed lines (crosses) show the contours (maxima) of the open case and 
solid lines (solid circles) show those of the flat-$\Lambda$ model. Panel 
$a)$ shows the $(\Omega_B h^2,\ \Omega_0)$ plane, and panel $b)$ shows the 
$(t_0, \ \Omega_0)$ plane.}
\end{figure}

\begin{figure}[p]
\psfig{file=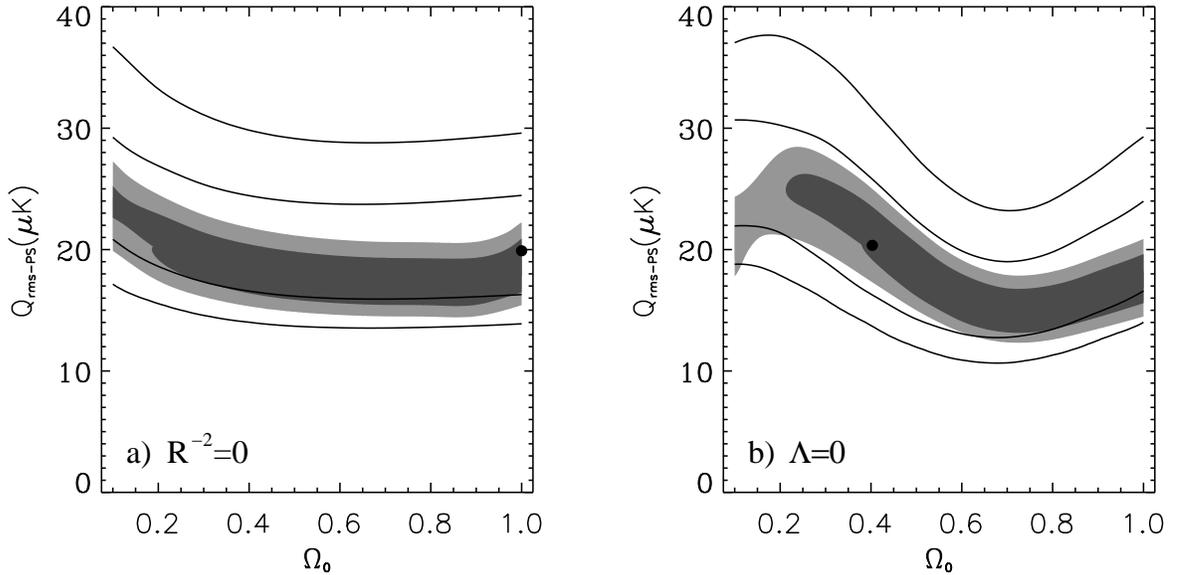,width=6.7in,angle=0}
\caption{Confidence contours and maxima of the two-dimensional 
$(Q_{\rm rms-PS}, \Omega_0)$ posterior probability density distribution 
functions. Panel $a)$ shows the flat-$\Lambda$ model and panel $b)$ 
the open model. Heavy lines show the $\pm1$ and $\pm2$ $\sigma$ confidence 
limits and solid circles show the maxima of the two-dimensional posterior 
distributions derived from the OVRO data. Shaded regions show the 
two-dimensional posterior distribution 1 $\sigma$ (denser shading) and 2 
$\sigma$ (less dense shading) confidence regions for the DMR data 
(G\'orski et al. 1998; Stompor 1997). The DMR results are a composite of 
those from analyses of the two extreme data sets: i) galactic frame with 
quadrupole included and correcting for faint high-latitude galactic 
emission; and ii) ecliptic frame with quadrupole excluded and no other 
galactic emission correction (G\'orski et al. 1998).}
\end{figure}

\begin{figure}[p]
\psfig{file=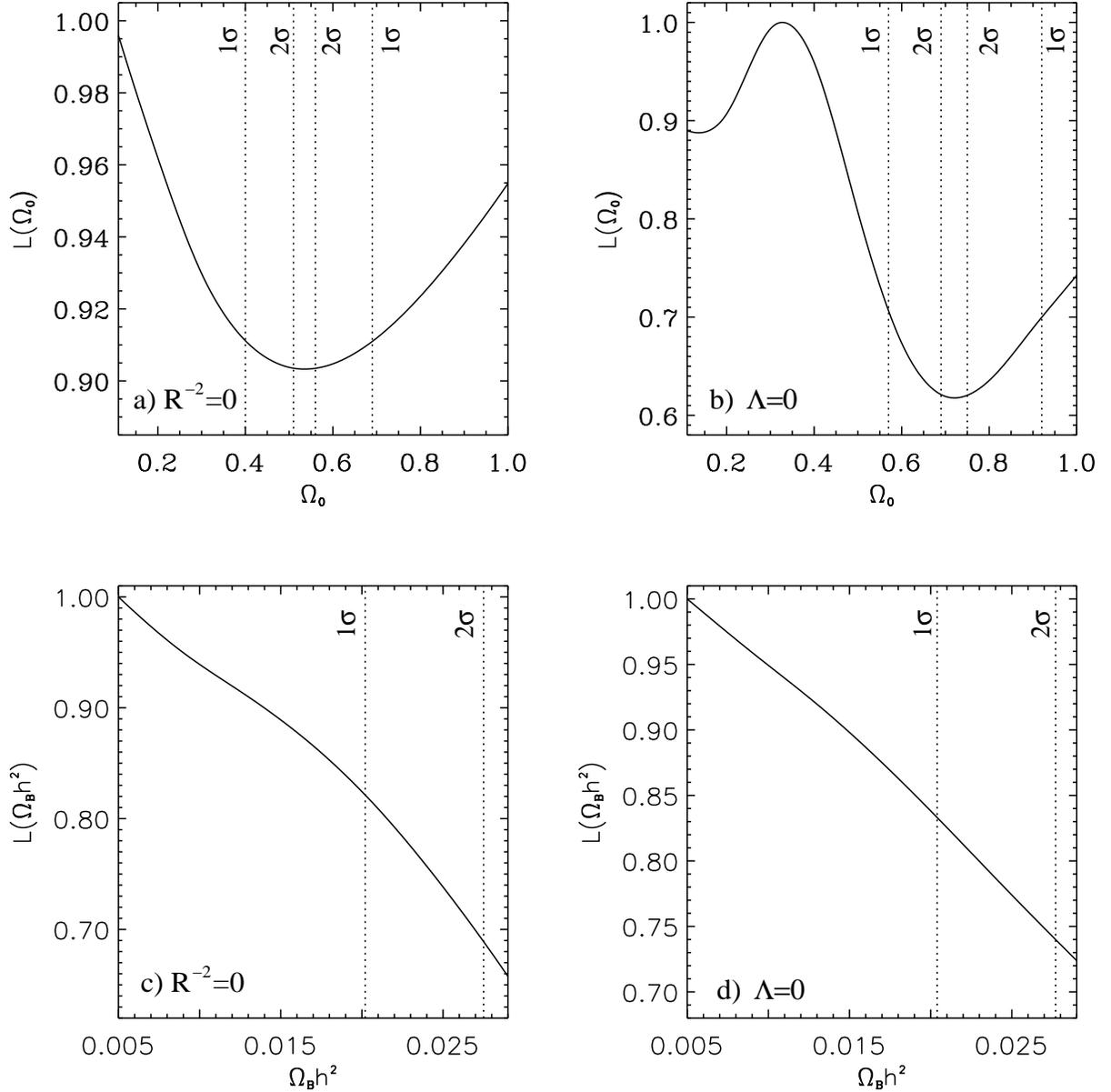,height=6.7in,width=6.7in,angle=0}
\caption{One-dimensional posterior probability density distribution 
functions for $\Omega_0$, $\Omega_B h^2$, $t_0$, and $Q_{\rm rms-PS}$ 
(derived by marginalizing the four-dimensional one over the other
three parameters) in the open and flat-$\Lambda$ models.  These have
been renormalized to unity at the peaks. Dotted vertical lines show the 
confidence limits derived from these one-dimensional posterior 
distributions and solid vertical lines in panels $g)$ and $h)$ show the 
$\pm 1$ and $\pm 2$ $\sigma$ confidence limits derived by projecting the 
OVRO data four-dimensional posterior distributions. The 2 $\sigma$ DMR 
(marginalized and projected) confidence limits in panels $g)$ and $h)$ 
are a composite of those from the two extreme DMR data sets (see caption 
of Fig. 2). When the four-dimensional posterior distributions are normalized 
such that $L(Q_{\rm rms-PS}\ =\ 0\ \mu{\rm K})\ =\ 1$, the peak values of
the one-dimensional distributions shown in panels $a)-h)$ are 
$1\times 10^{67}$, $2\times 10^{67}$, $6\times 10^{68}$, $7\times 10^{68}$, 
$1\times 10^{66}$, $2\times 10^{66}$, $1\times 10^{66}$, and 
$1\times 10^{66}$, respectively.}
\end{figure}

\begin{figure}[p]
\psfig{file=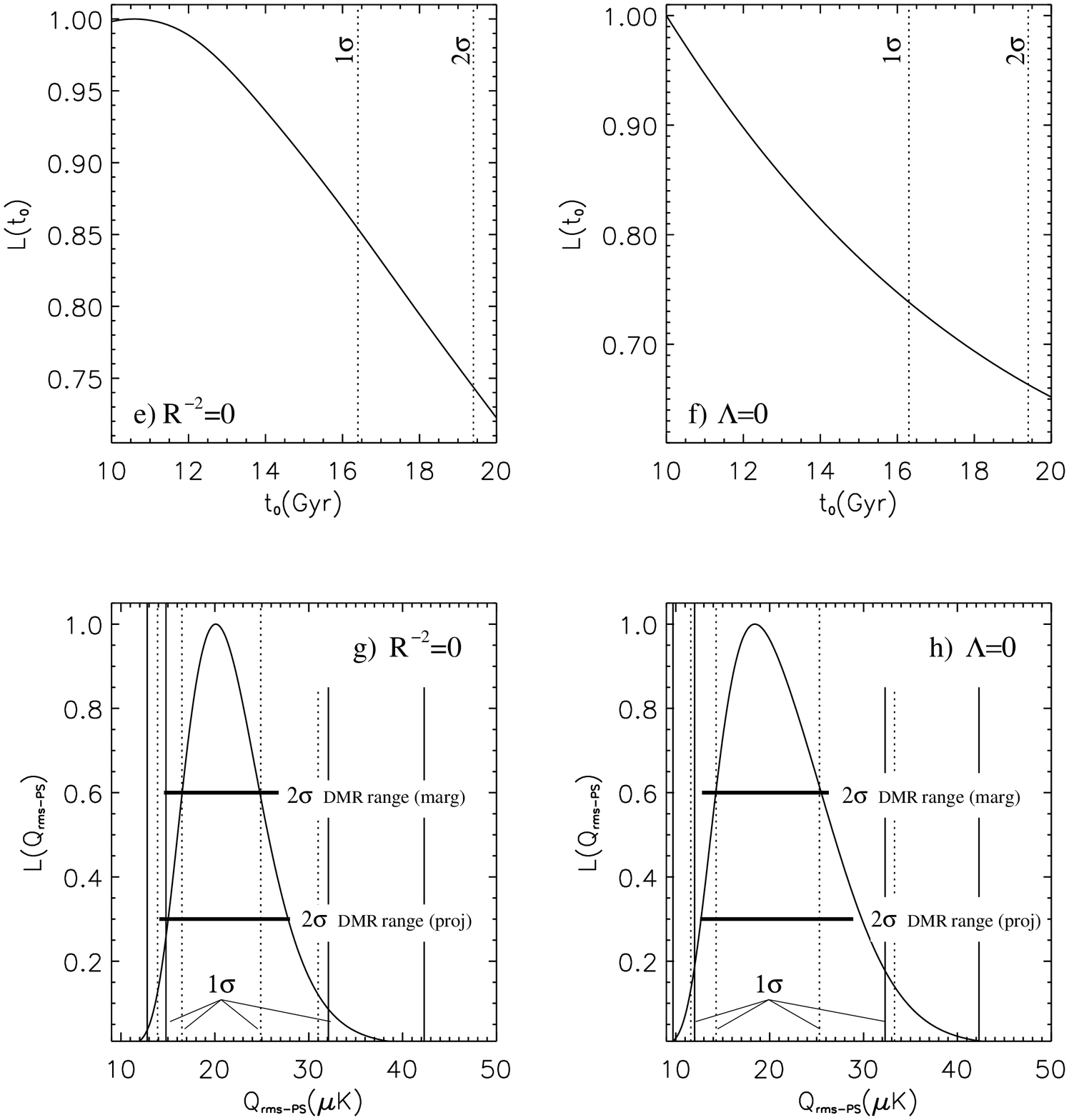,height=6.7in,width=6.7in,angle=0}
\end{figure}

\end{document}